\documentclass{article}
\usepackage{harvard}
\usepackage{amssymb}
\usepackage{amsmath}

\input{tcilatex}
\begin{document}

\title{\textbf{Towards Ghost-Free Gravity and Standard Model}}
\author{V. I. Tkach \\
Department of Physics and Astronomy,\\
Northwestern University, Evanston, IL 60208-3112, USA\\
\textit{v-tkach@northwestern.edu}}
\maketitle

\begin{abstract}
This paper presents a new higher derivative gravity which in spontaneous
breaking electroweak symmetry state does not have ghost in gravity sector.
We show that Newton constant of the gravity and dark energy density they
depend on the fundamental TeV scale and the coupling constant at the
quadratic curvature term. We consider the supersymmetric extension of this
model.

\end{abstract}

\begin{center}

\ \ 
\end{center}

It is well known that the Einstein action of general relativity leads to the
non-renormalizable quantum theory $^{\text{1}}$. The Einstein theory should
be a good approximation at classical level and has a sensible Newtonian
limit.\ \ \ \ \ \ 

The theory of the higher derivative gravitation, whose action contains terms
quadratic in the curvature in addition to the Einstein term, is a
renormalizable field theory $^{\text{2-8}}$, but it is not free of defects.
As such, these theories contain both second and fourth order derivative to
gravitational components. It gives rise to unphysical poles in spin two
sector of the tree- level propagator which breaks the unitarity. A possible
way to overcome this problem is to consider nonlocal gravity $^{\text{9-10}}$
and the idea proposed in $^{\text{11}}$ is to modify the ultraviolet
behavior of the graviton propagator in Lorenz non- invariant way.

On the other hand, induced gravity program $^{\text{12-15}}$ with fourth
-derivative gravitational theories do not contain dimensional coupling
constants and the unphysical ghost. However, in such theories Newton's
constant is not calculable and is a free parameter $^{\text{16}}$, and does
not have the Newtonian limit.

In the previous work $^{\text{17}}$ it was shown that the electroweak
symmetry breaking can be used for the construction of the quantum gravity
free of defects.

This paper is devoted to the investigation of an example of the quantum
gravity with higher curvature which is ghost-free.

We show that the gravitational strength, other observed fundamental
interactions, and vacuum energy density are the consequence of one
fundamental dimensional scale $\ M_{EW}\thicksim 2\times 10^{3}$GeV , which
depends on vacuum expectation value of the Higgs fields $\ <\varphi
^{0}>\simeq 250$GeV of the Standard Model : $v=M_{EW}=8<\varphi ^{0}>\simeq
2\times 10^{3}$GeV and the dimensionless coupling constant at quadratic
curvature term .

In this paper, we adopt the units $c=\hslash =1$.

Let us consider a new model coupling of the gravity to the Standard Model
with the uniquely formed action as follows

\begin{equation}
S=\dint d^{4}x\sqrt{-g}[-\dfrac{\epsilon }{2}\left( v^{2}-8^{2}\Phi ^{+}\Phi
\right) R+8\beta G_{\mu \rho }R^{\mu \rho }-
\end{equation}

\begin{equation*}
-\frac{1}{2}\left( D_{\mu }\Phi \right) ^{+}\left( D^{\mu }\Phi \right) -%
\frac{f}{8}\left( 8^{2}\Phi ^{+}\Phi -v^{2}\right) ^{2}]+S_{SM},
\end{equation*}

where $S_{SM}$ is a part of the action Standard Model for gauge fields and
the fermion fields.

The fundamental scalar doublet is $\Phi ^{T}=(\phi ^{+},\phi ^{0})$ of \
Higgs fields $^{\text{18}}$, and $G_{\mu \rho }=R_{\mu \rho }-\frac{1}{2}%
g_{\mu \rho }R$ is the Einstein tensor, where $R_{\mu \rho }$\ is the Ricci
tensor and $R$\ is the scalar curvature. In the action (1) $\epsilon $, $%
\beta $, and $f$ are dimensionless coupling constants. The action (1) is
perturbatively renormalizable $^{\text{2,5}}$, but has the ghost to spin two
sector and tachyon to Higgs sector in unbroken the electroweak symmetry. It
is a known fact that no new ultraviolet divergences occur in theory with
spontaneous symmetry breaking, over and above those in an unbroken theory $^{%
\text{19}}$. Hence, spontaneous breaking of the symmetry does not affect
renormalization.

We suggest that Einstein's action is be modified to read $-\frac{1}{2}%
\epsilon v^{2}R\sqrt{-g}$\ \ in the action (1), where we have rule $\epsilon
v^{2}=M_{p}^{2}$\ and $M_{p}=(8\pi G)^{-\frac{1}{2}}\simeq 2.4\times 10^{18}$%
GeV is the reduced Planck mass. So the Newton's constant $G$ is not the
fundamental constant and show that $\epsilon =\beta ^{\frac{1}{2}}$, where $%
\beta $ is the coupling constant of the higher curvature gravity.\qquad

\qquad The higher curvature term in the action (1) has little effect at low
energies by compared to the Einstein term. At the lowest energy, only $-%
\frac{1}{2}\epsilon v^{2}R\sqrt{-g}$ is important to the current
experimental \ \ tests of Newton's law $^{\text{20}}$ that does not
contradict with coupling constant $\beta $ which has the value of $\beta
\simeq 2\times 10^{60}$ . The current experimental constraints from
sub-millimeter tests to corrections of the higher curvature term $^{\text{%
21,22}}$ to the Newtonian potential , give for constant $\ \beta $ bounding $%
\beta <10^{62}$.

\qquad The field equations for metric $g_{\mu \sigma }$ and the Higgs fields 
$\Phi $ following from the action (1) have solutions $g_{\mu \sigma
}^{(0)}=\eta _{\mu \sigma }$\ is the Minkowski metric as the metrical ground
state and nontrivial \ \ Higgs fields ground state is $8^{2}\left( \Phi
^{+}\Phi \right) _{0}=v^{2}\approx (2\times 10^{3}$GeV$)^{2}$. This is
ground state with energy zero. The Higgs mechanism requires that the
unbroken state has $<\Phi >=0,$ and the vacuum broken state has $<\Phi
^{T}>=(0,\frac{v}{8})$.

\qquad The standard way in perturbative theory is to write the metric as $%
g_{\mu \sigma }=\eta _{\mu \sigma }+\widetilde{h}_{\mu \sigma }$. In the
unitarity gauge Higgs fields takes the following form, avoiding Goldstone
bosons

\begin{equation}
\ \Phi =%
\begin{pmatrix}
0 \\ 
\frac{v}{8}+\varphi%
\end{pmatrix}%
\ \ ,
\end{equation}

where the real scalar field $\varphi (x)$\ describes the excited Higgs field
connected with the Higgs particle. \ 

\bigskip We have that $W_{\mu }^{\pm }$ \ and \ Z$_{\mu }$ \ gauge bosons
pick up masses from the spontaneous breaking of the electroweak symmetry: $%
M_{W}^{2}=\frac{g^{2}v^{2}}{4\cdot 8^{2}}$, $M_{Z}^{2}=\frac{M_{W}^{2}}{\cos
^{2}\theta _{W}}$.

\ From the relation $\frac{g^{2}}{8M_{^{W}}^{2}}=\frac{4\cdot 8}{v^{2}}=%
\frac{G_{F}}{\sqrt{2}}$, where $G_{F}\simeq 1.17\times 10^{-5}$GeV$^{-2}$ is
Fermi constant from muon decay, we obtain $(\frac{v}{8})^{2}$\ $=\frac{1}{%
\sqrt{2}G_{F}}\simeq (250$GeV$)^{2}$.

Thus, after the spontaneous symmetry breaking we have the electroweak scale
\ $v\simeq 2\times 10^{3}$GeV fixed by the Fermi weak coupling constant $%
G_{F}$. \ \ \ 

The part of the action (1) quadratic in the fields $\widetilde{h}_{\mu
\sigma }$ and $\varphi $ in the state of the electroweak symmetry breaking
can be written as

\begin{equation}
\ S=\dint d^{4}x[8\epsilon v\varphi R^{(1)}(\widetilde{h})+8\beta G_{\mu
\rho }^{(1)}(\widetilde{h})R^{(1)\mu \rho }(\widetilde{h})-\frac{1}{2}%
\partial _{\mu }\varphi \partial ^{\mu }\varphi -\dfrac{8^{2}fv^{2}}{2}%
\varphi ^{2}]
\end{equation}

where the Ricci tensor $R_{\mu \rho }^{(1)}(\widetilde{h})$ and the scalar
curvature $R^{(1)}(\widetilde{h})$ can be written in a linearized form

\begin{equation}
R_{\mu \rho }^{(1)}(\widetilde{h})=\frac{1}{2}(\square \widetilde{h}_{\mu
\rho }-\partial _{\mu }\partial _{\sigma }\widetilde{h}_{\rho }^{\sigma
}-\partial _{\rho }\partial _{\sigma }\widetilde{h}_{\mu }^{\sigma
}+\partial _{\mu }\partial _{\rho }\widetilde{h}),\ \ 
\end{equation}

\begin{equation}
R^{(1)}(\widetilde{h})=(\square \widetilde{h}-\partial ^{\rho }\partial
^{\sigma }\widetilde{h}_{\rho \sigma })\ 
\end{equation}

where \ $\square =\partial _{\mu }\partial ^{\mu }$ denotes the flat
space-time d'Alamberian.

The expression (3) for the fields $\widetilde{h}_{\mu \rho }$ and $\varphi $
has the unwanted mixed term

\begin{equation*}
\ \ 8\epsilon v\varphi R^{(1)}(\widetilde{h}).
\end{equation*}%
\ \ \ \ \ \ \ \ \ \ \ \ \ \ \ \ \ \ \ \ \ \ \ \ \ \ \ \ \ \ \ \ \ \ \ \ \ \
\ \ \ \ \ \ \ \ \ \ \ \ \ \ \ \ \ \ \ \ \ \ \ \ \ \ \ \ \ \ \ \ \ \ \ \ \ \
\ \ \ \ \ \ \ \ \ \ \ \ \ \ 

We can get rid of this term making the following redefined field

\begin{equation}
\widetilde{h}_{\mu \rho }=h_{\mu \rho }+\dfrac{\eta _{\mu \rho }v}{\epsilon }%
\square ^{-1}\varphi
\end{equation}

where $\square ^{-1}$ is the Green's function of the usual d'Alamberian
action on the Higgs field $.$ \ \ \ \ \ \ 

We find that the terms $R^{(1)}(\widetilde{h})$ and $G_{\mu \rho }^{(1)}(%
\widetilde{h})R^{(1)\mu \rho }(\widetilde{h})$ take the forms

\begin{equation}
\ \ R^{(1)}(\widetilde{h})=R^{(1)}(h)+\dfrac{3v}{\epsilon }\varphi \ \ 
\end{equation}

and

\begin{equation}
G_{\mu \rho }^{(1)}(\widetilde{h})R^{(1)\mu \rho }(\widetilde{h})=G_{\mu
\rho }^{(1)}(h)R^{(1)\mu \rho }(h)-\frac{v}{\epsilon }\varphi R^{(1)}(h)-%
\frac{3v^{2}}{2\epsilon ^{2}}\varphi ^{2}
\end{equation}

we will not keep total derivative term in eq.(8).

Putting expressions (7) and (8) in the action (3) we get the following
condition rid of the mixed term

\begin{equation}
\ \ \epsilon ^{2}=\beta \ 
\end{equation}

for gravitational constants $\epsilon $ and $\beta $. \ \ \ \ \ \ \ \ \ \ \
\ \ \ \ \ \ \ \ \ \ \ \ \ \ \ \ \ \ \ \ \ \ \ \ \ \ \ \ \ \ \ \ \ \ \ \ \ \
\ \ \ \ \ \ \ \ \ \ \ \ \ \ \ \ \ \ \ \ \ \ \ \ \ \ \ \ \ \ \ \ \ \ \ \ \ \
\ \ \ \ \ \ \ \ \ \ \ \ \ \ \ \ \ \ \ \ \ \qquad

\qquad Therefore, the Planck scale $M_{p}$\ is not the fundamental scale and
depends

on the coupling constant $\beta \simeq 2\times 10^{60}$\ by quadratic
curvature term and

the electroweak scale $v\approx 2\times 10^{3}$GeV\ \ \ which\ \ is the
fundamental scale

\begin{equation}
\ \ M_{p}=(\beta )^{\frac{1}{4}}v\simeq 1.2\times 10^{15}\cdot 2\times
10^{3}GeV\simeq 2.4\times 10^{18}GeV.\ 
\end{equation}

As a result, expression (3) has the following form

\begin{equation}
\ S=\dint d^{4}x[8\beta G_{\mu \rho }^{(1)}(h)R^{(1)^{\mu \rho }}(h)-\frac{1%
}{2}\partial _{\mu }\varphi \partial ^{\mu }\varphi -\frac{1}{2}%
(8^{2}f-3\cdot 8)v^{2}\varphi ^{2}]\ \ \ 
\end{equation}

where $(8^{2}f-3\cdot 8)v^{2\text{\ }}=m_{\varphi }^{2}$ is the square mass
of the Higgs particle at $(8f-3)\geqslant 0$. The redefinition (6) \ does
not lead to appearance ghost in the sector Higgs particle, but leads to the
shift in square mass of the Higgs particle. The Higgs potential in the
Standart Model is unstable against quantum corrections.

A well known problem in physics is the existence of a huge gap between the
Standard Model scale and the Planck scale of gravity. The hierarchy problem $%
^{\text{23,24}}$ ,the \ stability of the Standard Model scale against the
Planck scale, is considered to be one of the most important issues in the
particle physics.

It has led to much of the original motivation for certain beyond the
Standard Model: low energy SUSY, little Higgs, gauge singlet scalars,
technicolor and so on.

In papers $^{\text{25,26}}$ it is shown that theories with a warped extra
dimensions and large extra spatial dimensions solve hierarchy problem
without supersymmetry or technicolor.

Our model automatically lowers the Planck scale cutoff to $M_{EW}=2\times
10^{3}$GeV ultraviolet (UV) cutoff, so it accounts for \ the quantum
stability of the Standard Model. It is a new solution of the hierarchy
problem.

\qquad Of course, the differential operator which appears in the gravity
part of action (11) is not invertible. It is necessary to add a gauge-
fixing term in this case

\begin{equation}
S_{GF}=-\dfrac{1}{2\alpha }\dint (\partial ^{\sigma }h_{\sigma \mu }\eta
^{\mu \lambda }\square \partial ^{\rho }h_{\rho \lambda })d^{4}x.\ \ \ 
\end{equation}

Going over to momentum space and using the projectors for the spin-two $%
P_{\mu \rho \lambda \sigma }^{(2)}$ , spin-one $P_{\mu \rho \lambda \sigma
}^{(1)}$ , the two spin-zero $P_{\mu \rho \lambda \sigma }^{(0-s)}$\ , and $%
P_{\mu \rho \lambda \sigma }^{(0-w)}$\ $^{\text{2,8,27}}$ we find for
actions (11) and (12)

\begin{equation}
\ \ \widetilde{S}=S+S_{GF}=\dfrac{1}{2}\dint h^{\mu \rho }(-k)\{k^{4}[4\beta
P^{(2)}+\dfrac{1}{2\alpha }P^{(1)}-8\beta P^{(0-s)}+\ \ \ 
\end{equation}

\begin{equation*}
+\dfrac{1}{\alpha }P^{(0-w)}]_{\mu \rho \lambda \sigma }\}h^{\lambda \sigma
}(k)d^{4}k.
\end{equation*}

\bigskip Then the propagator for the fields $h_{\lambda \rho \text{ }}$in
the momentum space is%
\begin{equation}
D_{\mu \rho \lambda \sigma }=\dfrac{2}{4\beta k^{4}}P_{\mu \rho \lambda
\sigma }^{(2)}+\dfrac{2\alpha }{k^{4}}P_{\mu \rho \lambda \sigma }^{(1)}-%
\dfrac{1}{8\beta k^{4}}P_{\mu \rho \lambda \sigma }^{(0-s)}+\dfrac{\alpha }{%
k^{4}}P_{\mu \rho \lambda \sigma }^{(0-w)}.
\end{equation}

The component projectors by $P^{(1)}$ and $P^{(0-w)}$\ can be gauged away

at $\ \alpha \rightarrow 0$.

Ignoring the terms proportional $\alpha $ in (14), we have the following
form for the propagator

\begin{equation}
\ \ D_{\mu \rho \lambda \sigma }=\dfrac{1}{4\beta k^{4}}(P_{\mu \rho \lambda
\sigma }^{(2)}-\dfrac{1}{2}P_{\mu \rho \lambda \sigma }^{(0-s)}).\ \ \ 
\end{equation}

Thus, \ $P^{(0-s)}$ residue is negative it is a ghost. There is the kind
ghost \ related to the $P^{(0-s)}$ projector which has precisely the
coefficient. It was actually \ necessary for the correct cancellation of the
unphysical longitudinal part of the $P^{(2)}$ projector. Thus we conclude
that the propagator describes the physical graviton state. If \ we have in
action (1) the terms \ $aW-\frac{b}{3}R^{2}$ (see ref.$^{\text{17}}$ ),

instead of the term $8\beta G_{\mu \rho }R^{\mu \rho }$, then the propagator
has the ghost to the $P^{(0-s)}$ projector and there is not cancellation of
the unphysical longitudinal part of the $P^{(2)}$ projector.

As the tree-level propagator (15) do not have the ghost that in a
consequence of the local Poincare symmetry loop corrections may still do not
destroy the ghost absence.

Let us note that redefinition (6) brings the contribution $\dfrac{v}{\beta ^{%
\frac{1}{2}}k^{2}}$ to some vertex of the Feynman diagrams. It leads us to
the necessity of introduction of an infrared (IR) cutoff at the Feynman
integrals.

We assume that the (IR) cutoff is $k_{IR}=L^{-1}=\frac{v}{\beta ^{3/4}}$ and
provides one identity the scale $L=\frac{\beta ^{3/4}}{v}$ with the horizon
size of the present

universe $L\sim \frac{1}{H}$ , where $H=1.3\times 10^{-42}$GeV is the Habble
parameter.

According to the holographic principle $^{\text{28,29}}$ the vacuum energy
density is $\rho _{vac}\sim 3d^{2}M_{p}^{2}L^{-2}$, where $d\lesssim 1$ is a
numerical parameter. The largest size $L$ compatible with this is the
infrared (IR) cutoff of the effective quantum field theory.

\qquad\ We have the following form of the vacuum energy density

\begin{equation}
\rho _{vac}\sim 3d^{2}M_{p}^{2}L^{-2}=3d^{2}\frac{v^{4}}{\beta }\sim
10^{-47}GeV^{4}\ \text{,}\ 
\end{equation}

at $\ \beta \simeq 2\times 10^{60}$ . Thus in the framework the new version $%
R^{2}$-gravity with one scale, which is the electroweak scale $v\simeq
2\times 10^{3}$GeV, can be also find of the solution smallness problem of
the cosmological constant. The Bekenstein-Hawking entropy for the present
universe is \ $S_{BH}\sim \pi M_{p}^{2}L^{2}=\pi \beta ^{2}\sim 10^{121}$.

Vacuum energy density (16) \ plays the role of the dark energy, which counts
about 75 percent of the total energy density. As a result, the universe
expansion is accelerating $^{\text{30}}$.

\ \ It has been known from many astrophysical measurements that the universe
contains about 20 percent of the total energy matter density, non baryonic

dark matter (DM) which is not included in the Standard Model (SM).

\ \ It has been studied $^{\text{31,32}}$ that a few multiplet ( gauge
singlets, doublet) that can be added to the SM without introducing
supersymmetry (SUSY)

as a potential dark matter candidate.

\ \ In this note, we make the SUSY extension of the action (1) which
includes the SUSY extension of the SM. The SUSY extension of the SM

to the Next - to Minimal Supersymmetric Standard Model (NMSSM) can have the
DM good scenarios. In order to obtain the corresponding N=1 \ SUSY

for the action (1), we followed the superfield approach ref.$^{\text{33,34}}$%
.

\ \ \ \ \ \ The supersymmetry generalization of the action (1) in chiral
superspace has the form

\begin{eqnarray}
S &=&\dint d^{4}x[d^{2}\theta 2E\{-3\epsilon (v^{2}-Y(\Phi ))R-\frac{1}{16}(%
\overline{D}^{2}-8R)\Phi ^{+}e^{V}\Phi - \\
&&-8\beta (\overline{D}^{2}-8R)(G_{m}G^{m}-\frac{1}{4}R^{+}R)+W(\Phi )+ 
\notag \\
&&+\text{gauge and quark, lepton superfields}\}+h.c.]\text{ .}  \notag
\end{eqnarray}

\bigskip The chiral superspace density or the vierbein multiplet (in
Wess-Zumino gauge) reads $2E(x,\theta )=e(x)[1+i\theta \sigma ^{a}\overline{%
\psi }_{a}-\theta ^{2}(M+\overline{\psi }_{a}\overline{\sigma }^{ab}%
\overline{\psi }_{b})]$, where

$e(x)=\sqrt{-\det g_{\mu \rho }}$ and $g_{\mu \rho }=e_{\mu }^{a}e_{a\rho }$
is the space-time metric, $\psi _{b}^{\alpha }(x)=e_{b}^{\mu }\psi _{\mu
}^{\alpha }$ is a gravitino, and $M(x)$ is the complex auxiliary field.

The gauge invariant function $Y(\Phi )$ is a quadratic function of the
chiral superfields $\Phi (x,\theta )$ which are the Higgs and the gauge
scalars supermultiplets.

The interactions of the $SU(2)_{L}$ and $U(1)_{Y}$ gauge field
supermultiplets $V(x,\theta ,\overline{\theta })$ with the Higgs
supermultiplets are represented by the factor $e^{V}$

with $V$ in the appropriate representation.

The chiral complex scalar superfield $R(x,\theta )$ is the curvature
supermultiplet containing the scalar curvature $R(x)$ at its $\theta ^{2}$
term.

The real vector superfield $G_{m}(x,\theta ,\overline{\theta })$ has the
traceless part of the Ricci tensor, $R_{\mu \nu }-\frac{1}{4}g_{\mu \nu }R$
in its $\theta \sigma ^{b}\overline{\theta }$ component.

The superpotential $W(\Phi )$ in case NMSSM depends exlusively on the Higgs
chiral superfields $H_{1}(x,\theta )$, $H_{2}(x,\theta )$ , and complex
scalar superfield $S(x,\theta )$

can be form%
\begin{equation}
W(\Phi )=\xi (H_{1}H_{2}-v^{2})S+\frac{\lambda S^{3}}{3}\text{ .}
\end{equation}

However to (17) \ we have to add the Yukawa couplings of the quarks and
leptons superfields.

The spontaneous supersymmetry breaking of the higher derivative supergravity
theory in the case $\phi (R(x,\theta ))$ it was consider in article $^{\text{%
35}}$.

The spontaneous supersymmetry breaking for our model (17) \ needs further
study. On the other hand, the vacuum energy density (16)

can be associated with SUSY breaking.

\bigskip

\textbf{References}

1. \ G. 't Hooft and M. Veltman, Ann. Inst. H. Poincare \textbf{20}, 69
(1974).

2. \ K. S. Stelle, Phys. Rev. D \textbf{16}, 953 (1977).

3. \ J. Julve and M. Tonin, Nuovo Cim. B \textbf{46}, 137 (1978).

4. \ A. Salam and J. Strathdee, Phys. Rev. D\textbf{\ 18}, 4480\textbf{\ }%
(1978).

5. \ I. G. Avramidi and A.O. Barvinsky, Phys. Lett. B \textbf{159}, 269
(1985).

6. \ E. Fradkin and A. A.Tseytlin, Nucl. Phys. B \textbf{201}, 469 (1982).

7.\ \ A. Codello and R. Percacci, Phys. Rev. Lett. \textbf{97}, 221301
(2006).

8. \ I. L. Buchbinder, S. D. Odintsov, and I. L. Shapiro,

\ \ \ \ \ \ Effective Action in Quantum Gravity,

\ \ \ \ \ \ IOP Publishing, Bristol and Philadelphia, 1992.

9. \ L. Modesto, arXiv :1107.2403 [hep-th].

10. \ T. Biswas, E. Gerwick, T. Koivisto, and A. Mazumdar,

\ \ \ \ \ \ \ \ \ Phys Rev. Lett. \textbf{108}, 031101 (2012), arXiv
:1110.5249 [gr-qc].

11. \ P. Ho\v{r}ava, Phys. Rev. D \textbf{79}, 084008 (2009), arXiv :
0901.3775 [hep-h].

12. \ S. Adler, Rev. Mod. Phys. \textbf{54}, 729 (1982) .

13. \ A. Zee, Ann. Phys. \textbf{151}, 431 (1983).

14. \ B.Hasslasher and F. Mottola, Phys. Lett. B \textbf{95}, 237 (1980).

15. \ I. L. Buchbinder and S. D. Odintsov,

\ \ \ \ \ \ \ \ \ \ \ \ \ \ Class. Quantum Grav. \textbf{2}, 721\textbf{\ }%
(1985).

16. \ F. David and A. Strominger, Phys. Lett. B \textbf{143}, 125 (1984).

17. \ V.I. Tkach, Mod. Phys. Lett. A\textbf{\ 24}, 1193 (2009),

\ \ \ \ \ \ \ \ \ \ \ \ arXiv : 0808.3429 [hep-th].

18. \ S. Weinberg, Phys. Rev. Lett. \textbf{19}, 1264\textbf{\ }(1967).

19. \ S. Coleman, S.Weinberg, Phys. Rev. D \textbf{7}, 1883 (1973).

20. \ D. J. Kapner, T. S Cook, E.G. Adelberger, J. H.Gundlalach,

\ \ \ \ \ \ \ \ B. R. Heckel, C. D. Joyle, and H. E. Swanson,

\ \ \ \ \ \ \ \ Phys. Rev.Lett. \textbf{98}, 021101 (2007).

21. \ J. F. Donoghue, Phys. Rev. D\textbf{\ 50}, 3874 (1994).

22. \ K. S. Stelle, Gen. Rel.Grav. \textbf{9}, 353 (1978).

23. \ L. Susskind, Phys. Rev. D \textbf{20}, 2619 (1979).

24. \ S. Weinberg, Phys. Rev. D \textbf{19}, 1277 (1979) .

25. \ L. Randall and R. Sundrum, Phys. Rev. Lett. \textbf{83}, 3370 (1999),

\ \ \ \ \ \ \ \ \ arXiv : hep- th / 9905221.

26. \ N. Arkani-Hamed, S. Dimopoulos and G. R. Dvali,

\ \ \ \ \ \ \ \ Phys. Lett. B\textbf{\ 429}, 263 (1998), arXiv : hep-th /
9803315.

27. \ P. van Nieuwenhuizen, Nucl. Phys. B \textbf{60}, 478 (1973).

28. \ S.D.H. Hsu, Phys. Lett. B\textbf{\ 594}, 13 (2004), arXiv : hep-th /
0403052.

29. \ M. Li, Phys. Lett. B \textbf{603}, 1 (2004), arXiv: hep-th/0403127.

30. \ A. G. Riess \textit{et al}., Astrophysics J. \textbf{607}, 665 (2004).

31. \ M. Cirelli, N. Fornego, and A. Strumia, Nucl. Phys. B \textbf{753},
178 (2006),

\ \ \ \ \ \ \ \ \ \ \ arXiv: hep-ph/ 0512090.

32. \ M. Gonderinger, H. Lim, and M. J. Ramsey-Musolf, arXiv :1202.1316.

33. \ J.Wess and J. Bagger, Supersymmetry and Supergravity,

\ \ \ \ \ \ \ \ \ Princeton University Press, 1992.

34. \ S. J. Gates, Jr., M. T. Grisaru, M. Ro\v{c}ek, and W. Siegel,

\ \ \ \ \ \ \ \ \ \ Superspace or 1001 Lessons in Supersymmetry,

\ \ \ \ \ \ \ \ \ \ Benjamin-Cummings Publ. Company, 1983.

35. \ A. Hindawi, B. A. Ovrut, and D. Waldram,

\ \ \ \ \ \ \ \ \ \ \ \ Phys. Lett. B \textbf{381}, 154 (1996),\ arXiv:
hep-th/ 9602075.

\bigskip

\bigskip

\end{document}